\documentclass[aps,prl,letter,twocolumn,showpacs,floats,superscriptaddress]{revtex4}
\usepackage{graphicx}
\usepackage{bm}
\usepackage{amsmath}
\usepackage{amsfonts}
\usepackage{amssymb}

\DeclareMathOperator{\cosech}{cosech}

\begin{document}

\title{Spatial nonlocal pair correlations in a repulsive 1D Bose gas}
\author{A. G. Sykes}
\affiliation{ARC Centre of Excellence for Quantum-Atom Optics, School of Physical
Sciences, University of Queensland, Brisbane, Queensland 4072, Australia }
\author{D. M. Gangardt}
\affiliation{School of Physics and Astronomy, University of Birmingham, Edgbaston,
Birmingham, B15 2TT, UK}
\author{M. J. Davis}
\affiliation{ARC Centre of Excellence for Quantum-Atom Optics, School of Physical
Sciences, University of Queensland, Brisbane, Queensland 4072, Australia }
\author{K. Viering}
\affiliation{Center for Nonlinear Dynamics, The University of Texas, Austin, Texas
78712-1081}
\author{M. G. Raizen}
\affiliation{Center for Nonlinear Dynamics, The University of Texas, Austin, Texas
78712-1081}
\author{K. V. Kheruntsyan}
\affiliation{ARC Centre of Excellence for Quantum-Atom Optics, School of Physical
Sciences, University of Queensland, Brisbane, Queensland 4072, Australia }
\date{\today}

\begin{abstract}
We analytically calculate the spatial nonlocal pair correlation function for
an interacting uniform 1D Bose gas at finite temperature and propose an
experimental method to measure nonlocal correlations. Our results span six
different physical realms, including the weakly and strongly interacting
regimes. We show explicitly that the characteristic correlation lengths are
given by one of four length scales: the thermal de Broglie wavelength, the
mean interparticle separation, the healing length, or the phase coherence
length. In all regimes, we identify the profound role of interactions and
find that under certain conditions the pair correlation may develop a global
maximum at a finite interparticle separation due to the competition between
repulsive interactions and thermal effects.
\end{abstract}

\maketitle

The study of two-body and higher-order correlations is becoming an important
theme in the physics of ultracold quantum gases. Correlation functions are
observables that provide information about quantum many-body wave functions
beyond the simple measurement of density profiles. They are of particular
importance for the understanding of low dimensional and strongly correlated
systems, atomic gases with exotic phases, and systems with multiple order
parameters. Such correlations can manifest themselves in momentum space, as
with pair correlations of a Fermi gas \cite{altman-demler-lukin}, and can be
observed in time-of-flight experiments \cite%
{Bloch-g2-MI,Greiner-dissociation} which has been the standard method of
measurement of degenerate gases. Nonlocal correlations in position, on the
other hand, should ideally be performed \textit{in-situ} with a spatial
resolution that is smaller than the typical correlation length, requiring
spatially resolved \textit{in-situ} single atom counting
\cite{Raizen-N-squeezing,Meschede,Grangier}.

In this paper we address the problem of nonlocal two-particle correlations
in a 1D Bose gas and propose an experimental method to measure them \textit{%
in-situ} using spatially resolved Raman transitions \cite%
{thomas-spatially-resolved} and single-atom counting in an optical box trap
\cite{Raizen-BEC-box}. We calculate the spatial second-order correlation
function $g^{(2)}(r)=\langle \hat{\Psi}^{\dagger }(0)\hat{\Psi}^{\dagger }(r)
\hat{\Psi}(r)\hat{\Psi}(0)\rangle /n^{2}$ for a uniform gas with repulsive $
\delta $-function interactions \cite{liebliniger,yangyang1} (where $\hat{\Psi
}$ is the field operator, $r$ is the interparticle separation, and $n$ is
the 1D linear density) and assess the feasibility of the method with respect
to the characteristic correlation lengths.

Given a particle at a certain location, the pair correlation $g^{(2)}(r)$
describes the probability of finding a second particle at a distance $r$
compared to uncorrelated particles, and gives the characteristic length
scale over which the density-density fluctuations decay. The knowledge of $%
g^{(2)}(r)$ in 1D Bose gases is of fundamental importance for the
understanding of second-order coherence and for practical applications such
as intensity interferometry in 1D environments. The recent experimental
realizations of ultracold atomic gases in the 1D regime \cite%
{1D-exp,Phillips-g3-1D-2003,Weiss-g2-1D,Schmiedmayer,vanDruten} and the fact
that the 1D Bose gas problem is exactly integrable using the Bethe ansatz
\cite{liebliniger,yangyang1} make this an ideal system for investigating
quantum many-body physics in previously unattainable regimes.

An early experimental measurement of atom-atom correlations using
microchannel plate detectors was performed in an ultracold (but not
degenerate) cloud of metastable neon \cite{Yasuda-Shimizu-HBT}. More
recently, the method was applied to quantum degenerate samples of helium
atoms and to correlations resulting from condensate collisions \cite%
{Orsay-3-exps}. Other experimental techniques to access higher-order
correlations include shot-noise spectroscopy of absorption images \cite%
{Bloch-g2-MI,Greiner-dissociation,Bouchoule-density-density}%
, the measurement of three-body recombination and photoassociation rates
\cite{Phillips-g3-1997,Phillips-g3-1D-2003,Weiss-g2-1D}, fluorescence
imaging \cite{Raizen-N-squeezing}, and atom counting using high-finesse
optical cavities \cite{Esslinger-g2}.

Certain aspects of pair correlations in a repulsive 1D Bose gas have
been studied previously, including the local correlation
$g^{(2)}(0)$, asymptotic properties at large $r$, and the zero
temperature behavior
\cite{girardeau,korepin-book,Castin,gangardt-correlations,karen-prl,karen-pra,drummond-canonical-gauge,1D-recent-theory,Cherny-Brand}.
Here we extend these results to nonlocal correlations in six
analytically tractable regimes of Ref. \cite{karen-prl,karen-pra},
ranging from strong to weak interactions -- all at finite
temperatures.

We begin by recalling that the second quantized Hamiltonian of the system is
given by
\begin{equation}
\hat{H}=\frac{\hbar ^{2}}{2m}\int dx\, \partial _{x} \hat{\Psi}^{\dagger
}\partial _{x}\hat{\Psi} +\frac{g}{2}\int dx\, \hat{\Psi}^{\dagger }\hat{\Psi%
}^{\dagger }\hat{\Psi} \hat{\Psi},
\end{equation}
where $m$ is the mass and $g>0$ is the coupling constant that can be
expressed via the 3D $s$-wave scattering length $a$ as $g\simeq
2\hbar ^{2}a/(ml_{\perp }^{2})=2\hbar \omega _{\perp }a$
\cite{olshanii-1d-scattering}. Here, we have assumed that the atoms
are transversely confined by a tight harmonic trap with frequency
$\omega_{\perp }$ and that $a$ is much smaller than the transverse
harmonic oscillator length $l_{\perp }=\sqrt{\hbar
/m\omega _{\perp }}$. The 1D regime is realized when the excitation energy $%
\hbar \omega _{\perp}$ is much larger than the thermal energy $T$ (with $%
k_{B}=1$) and the chemical potential $\mu$ \cite%
{karen-pra,interaction-crossover}. A uniform system in the thermodynamic
limit is completely characterized \cite{liebliniger,yangyang1} by two
parameters: the dimensionless interaction strength $\gamma =mg/(\hbar ^{2}n)$
and the reduced temperature $\tau =T/T_{d}$ \cite{karen-prl}, where $%
T_{d}=\hbar ^{2}n^{2}/(2m)$ is the temperature of quantum degeneracy.

Although the uniform 1D Bose gas problem is exactly solvable by the Bethe
ansatz \cite{liebliniger}, the cumbersome nature of the eigenstates
restricts the straightforward calculation of correlation functions~\cite%
{korepin-book,Sutherland}. The Hellmann-Feynman theorem and the solutions to
Lieb-Liniger \cite{liebliniger} or Yang-Yang \cite{yangyang1} integral
equations -- used for calculating the local correlation $g^{(2)}(0)$ \cite%
{gangardt-correlations,karen-prl,karen-pra} -- can no longer be applied to $%
g^{(2)}(r)$ at arbitrary separation $r$. For sufficiently large $r$, the
Luttinger liquid theory predicts universal features of $g^{(2)}(r)$ \cite%
{Giamarchi-book}. At $T=0$ its behavior is characterized by
interaction-dependent power-law approach to the uncorrelated value $%
g^{(2)}(r)=1$ at $r\rightarrow \infty $. At finite $T$ the approach to $%
g^{(2)}(r)=1$ becomes exponential. The Luttinger picture is limited to
temperatures smaller than the high energy cut-off of the order of $T_{d}$.
To describe non-universal features of $g^{(2)}(r)$ at high temperatures and
short-distances it is necessary to adopt alternative theoretical techniques
\cite{gangardt-correlations,karen-prl,g2r-epaps}.

\textit{Strongly interacting regime }[$\gamma \gg \max (1,\sqrt{\tau })$].
-- We employ perturbation theory with respect to the small parameter $\gamma
^{-1}$ \cite{Cheon-Shigevare} around the Tonks-Girardeau (TG) limit of
impenetrable (hard-core) bosons \cite{girardeau}. At $T=0$ we obtain \cite%
{g2r-epaps} the known \cite{korepin-book,Cherny-Brand} result
\begin{eqnarray}
g_{T=0}^{(2)}(r) &=&1-\frac{\sin ^{2}(z)}{z^{2}}-\frac{4}{\gamma }\frac{\sin
^{2}(z)}{z^{2}}-\frac{2\pi }{\gamma }\frac{\partial }{\partial z}\frac{\sin
^{2}(z)}{z^{2}}  \notag \\
&&+\frac{2}{\gamma }\frac{\partial }{\partial z}\left[ \frac{\sin (z)}{z}%
\int\nolimits_{-1}^{1}dt\sin (zt)\ln \frac{1+t}{1-t}\right],  \label{TGa}
\end{eqnarray}%
where $z=\pi nr$. The last term here diverges logarithmically with $z$ and
can be regarded as a first order perturbation correction to the fermionic
inverse square power law. Accordingly, Eq.~(\ref{TGa}) is valid for $z \ll
\exp(\gamma ^{-1})$.

Well below quantum degeneracy, $\tau \ll 1$, finite temperature corrections
are obtained using a Sommerfeld expansion around Fermi-Dirac distribution
for quasi-momenta at $T=0$. For $rn\ll \tau ^{-1}$ this gives \cite%
{g2r-epaps} an additional contribution $\tau ^{2}\sin ^{2}(\pi nr)/12\pi
^{2} $ to the rhs of Eq.~(\ref{TGa}), which is negligible compared to the $%
T=0$ result as $\tau \ll 1$. At $r=0$, Eq.~(\ref{TGa}) gives perfect
antibunching $g^{(2)}(0)=0$, which corresponds to a fully
\textquotedblleft fermionized\textquotedblright\ 1D Bose gas, where
the strong interatomic repulsion mimics the Pauli exclusion
principle for intrinsic fermions. By extending the perturbation
theory to include terms of order $\gamma ^{-2}$
we reproduce the known $T=0$ result for the local correlation, $%
g^{(2)}(0)=4\pi ^{2}/3\gamma ^{2}$ \cite{karen-prl,gangardt-correlations}.

\begin{figure}[tbp]
\includegraphics[height=3.8cm]{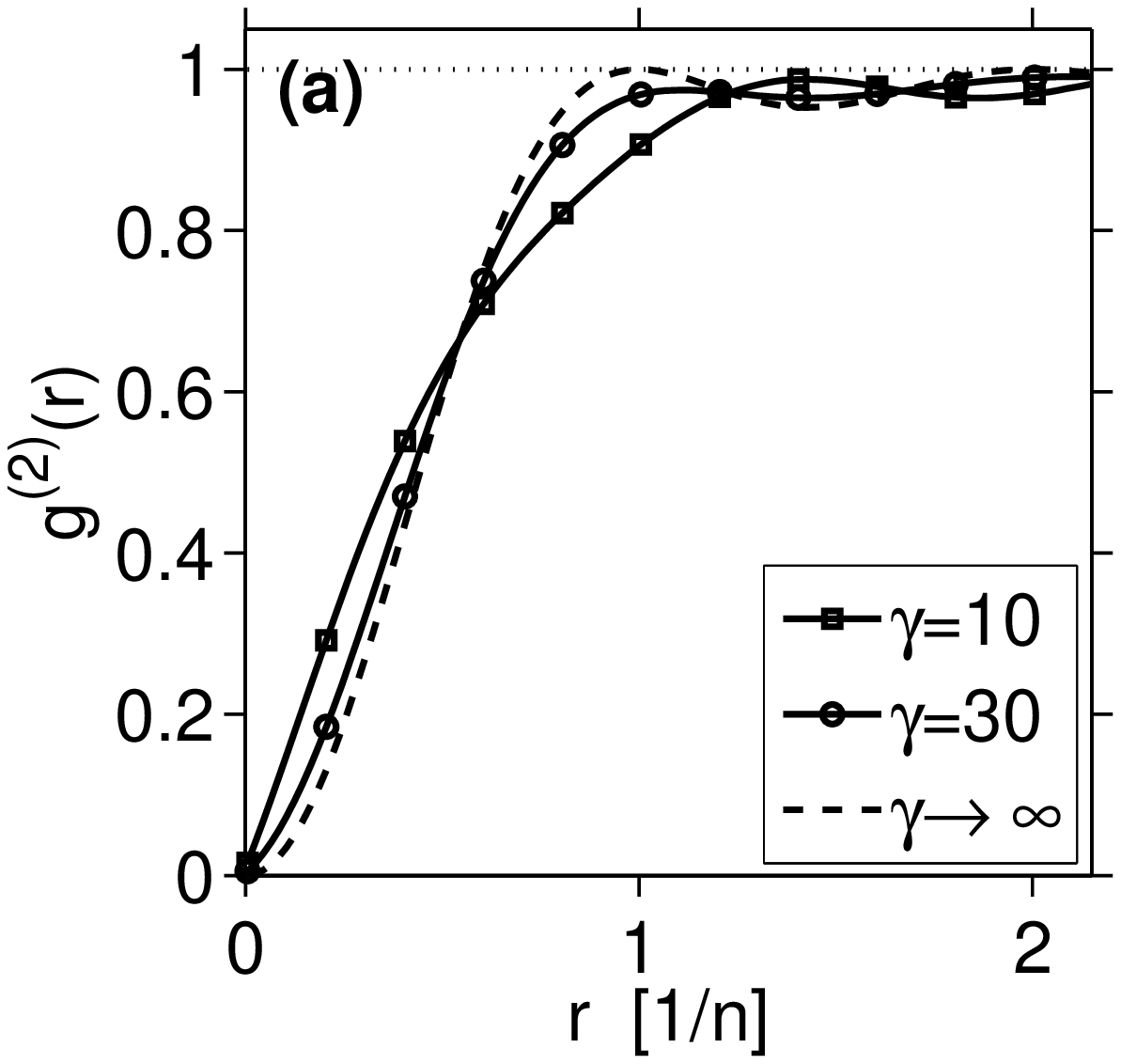} %
\includegraphics[height=3.8cm]{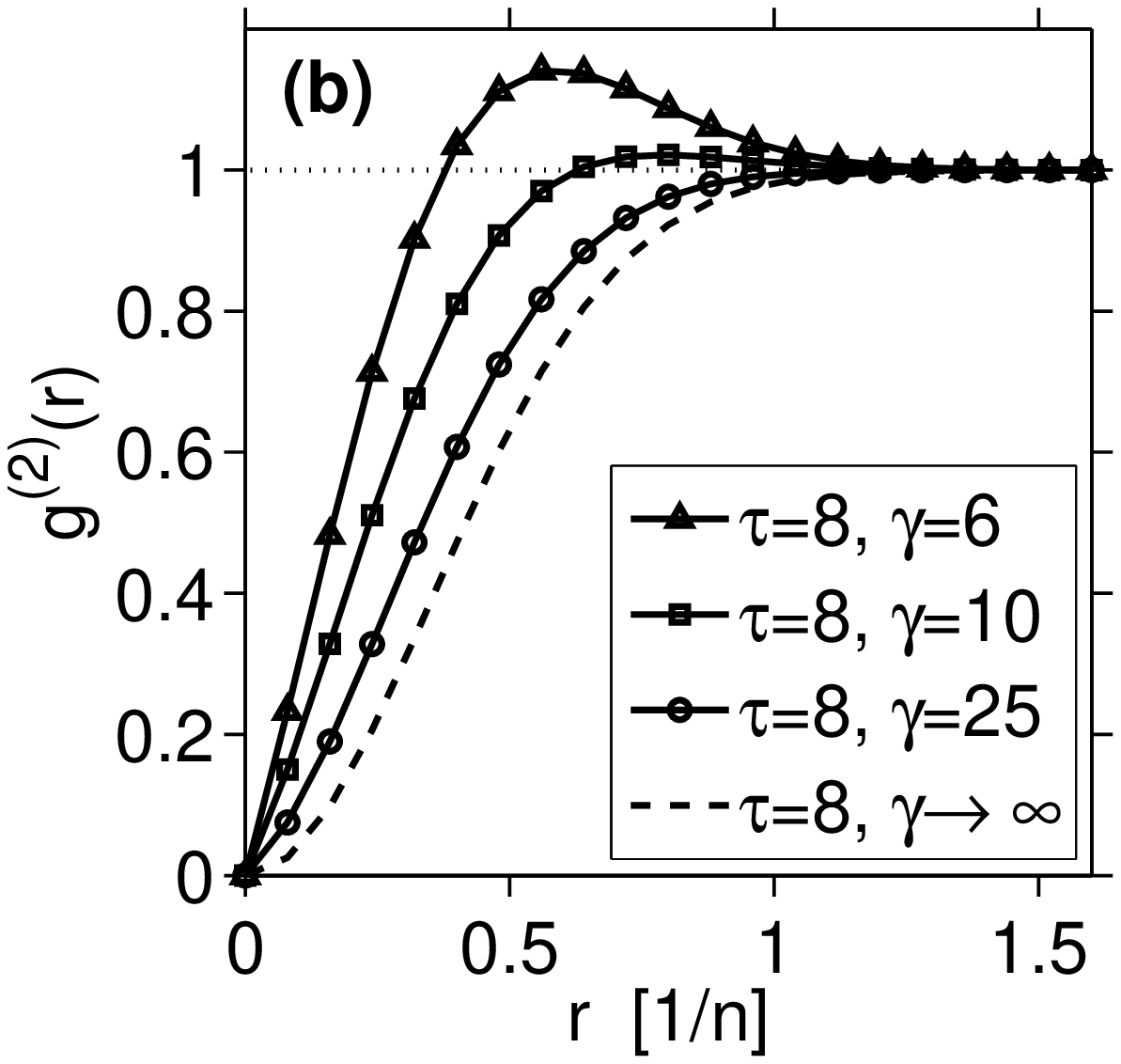}
\caption{Pair correlation $g^{(2)}(r)$ as a function of the relative
distance $r$ (in units of $1/n$) in the strongly interacting regime, $%
\protect\gamma \gg 1$: (a) low temperature TG regime, Eq.~(\protect\ref{TGa}%
), for $\protect\tau=0.01$; (b) regime of high-temperature \textquotedblleft
fermionization\textquotedblright, Eq.~(\protect\ref{TGb}).}
\label{TGfig}
\end{figure}

In Fig.~\ref{TGfig}(a) we plot the function $g^{(2)}(r)$, Eq.~(\ref{TGa}),
for various $\gamma $. The pair correlation exhibits oscillatory structure
(similar to Friedel oscillations of a 1D electron gas with an impurity \cite%
{Friedel}), with the local maxima implying the existence of more
likely separations between particles. Despite uniform density, this
can be interpreted as a quasi-crystalline order (with a period $\sim
1/n$) in the two-particle sector of the many-body wave function.


Well above quantum degeneracy, $\tau \gg 1$, we use perturbation theory
combined with the Maxwell-Boltzmann distribution of quasi-momenta. In this
high-temperature \textquotedblleft fermionization\textquotedblright\ regime
\cite{karen-prl} the characteristic momentum of the particles is $1/\Lambda
_{T}$ and the perturbation parameter is $a_{1D}/\Lambda _{T}\ll 1$, where $%
a_{1D}=\hbar ^{2}/mg\simeq l_{\perp }^{2}/a$ $\sim 1/\gamma n$ is the 1D
scattering length and $\Lambda _{T}=\sqrt{2\pi \hbar ^{2}/mT}$ is the
thermal de Broglie wavelength. This implies $\tau \ll \gamma ^{2}$, and to
first order in $\gamma ^{-1}$ we find \cite{g2r-epaps}
\begin{equation}
g^{(2)}(r)=1-(1-2\tau nr/\gamma )\;e^{-\tau n^{2}r^{2}/2}.  \label{TGb}
\end{equation}%
In the limit $r\rightarrow 0$ this leads to perfect antibunching, $%
g^{(2)}(0)=0$, while the corrections [as in Ref.~\cite{karen-prl}, $%
g^{(2)}(0)=2\tau /\gamma ^{2}$] are reproduced at second order in $\gamma
^{-1}$. The correlation length associated with the Gaussian decay is given
by $\Lambda _{T}=\sqrt{4\pi /(\tau n^{2})}$. For not very large $\gamma $,
the correlations show a nonmonotonic behavior with a global maximum at $%
r_{\max }\simeq \gamma /2\tau n$. This originates from the competition
between the interaction induced repulsion at short range and thermal
bunching [$g^{(2)}(r)>1$] at $r\sim \Lambda _{T}$. As $\gamma $ is increased
the position of the maximum diverges and its value approaches $1$ in a
non-analytical fashion $g^{(2)}(r_{\max })\simeq 1+(4\tau /\gamma ^{2})\exp
(-\gamma ^{2}/8\tau )$.

Figure~\ref{TGfig}(b) shows a plot of Eq.~(\ref{TGb}) for various $\gamma $
and $\tau $. For a well-pronounced global maximum, moderate values of $%
\gamma ^{2}/\tau $ are required (such as $\gamma ^{2}/\tau \simeq 5$, with $%
\tau =8$, $\gamma =6$), and these lie near the boundary of validity of
our
approximations ($\gamma ^{2}/\tau \gg 1$).
The exact
numerical calculations of Ref.~\cite{drummond-canonical-gauge}
provide further support for this result and show a similar maximum for $%
\gamma ^{2}/\tau \simeq 0.25$ (with $\tau =4\pi \times 10$ and
$\gamma =10$).

\textit{Weakly interacting regime} [$\tau^{2}\ll \gamma \ll 1$]. ---
For weak interactions we rely on the fact that the equilibrium state
of the gas is a quasi-condensate with suppressed density fluctuations and a
fluctuating phase ~\cite{mermin-wagner-hohenberg,petrov-1d-regimes}. The
pair correlation function is close to one and the deviations can be
calculated
using
Bogoliubov theory \cite{karen-prl,gangardt-correlations}.

\begin{figure}[tbp]
\includegraphics[height=3.8cm]{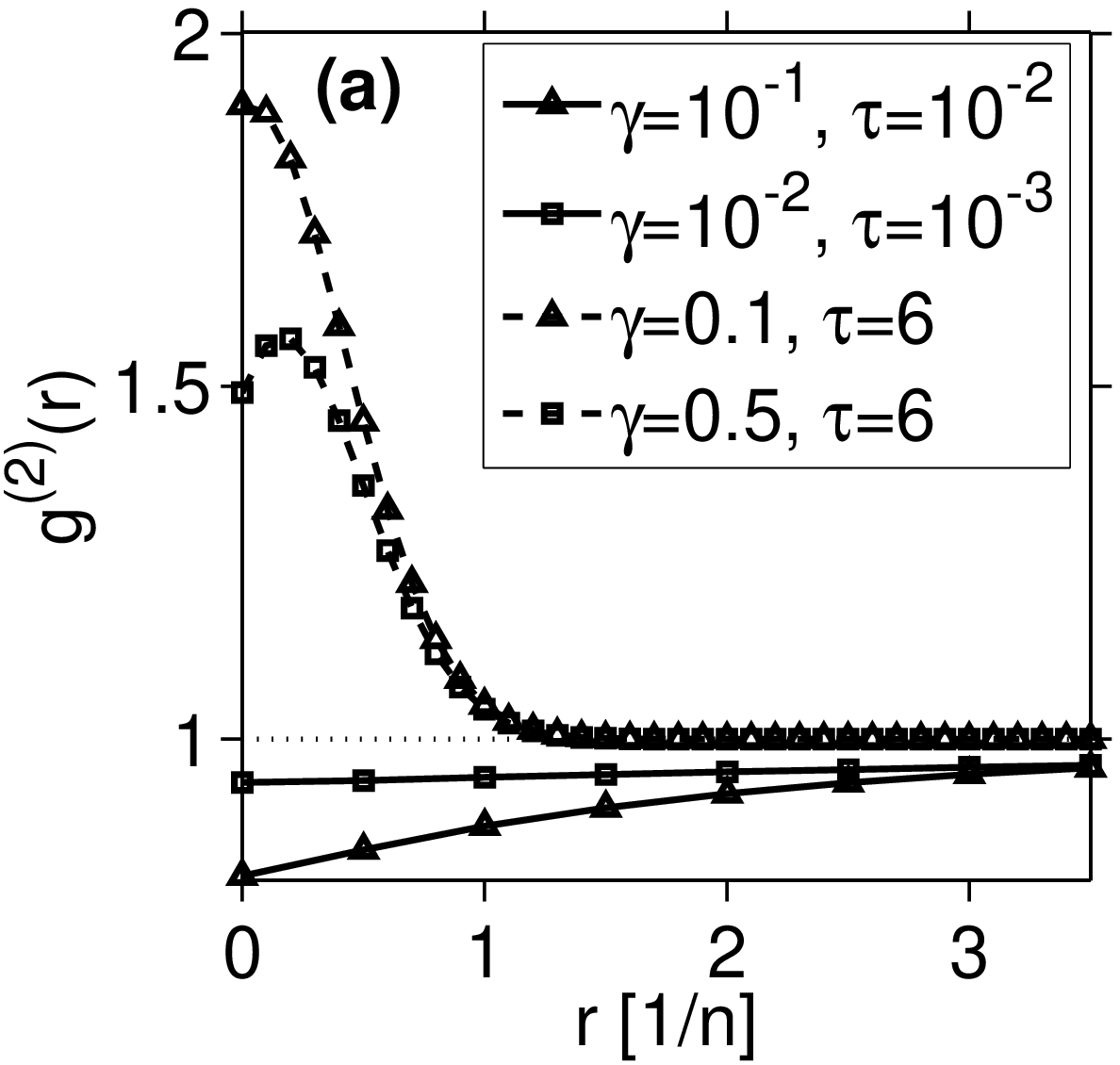} %
\includegraphics[height=3.8cm]{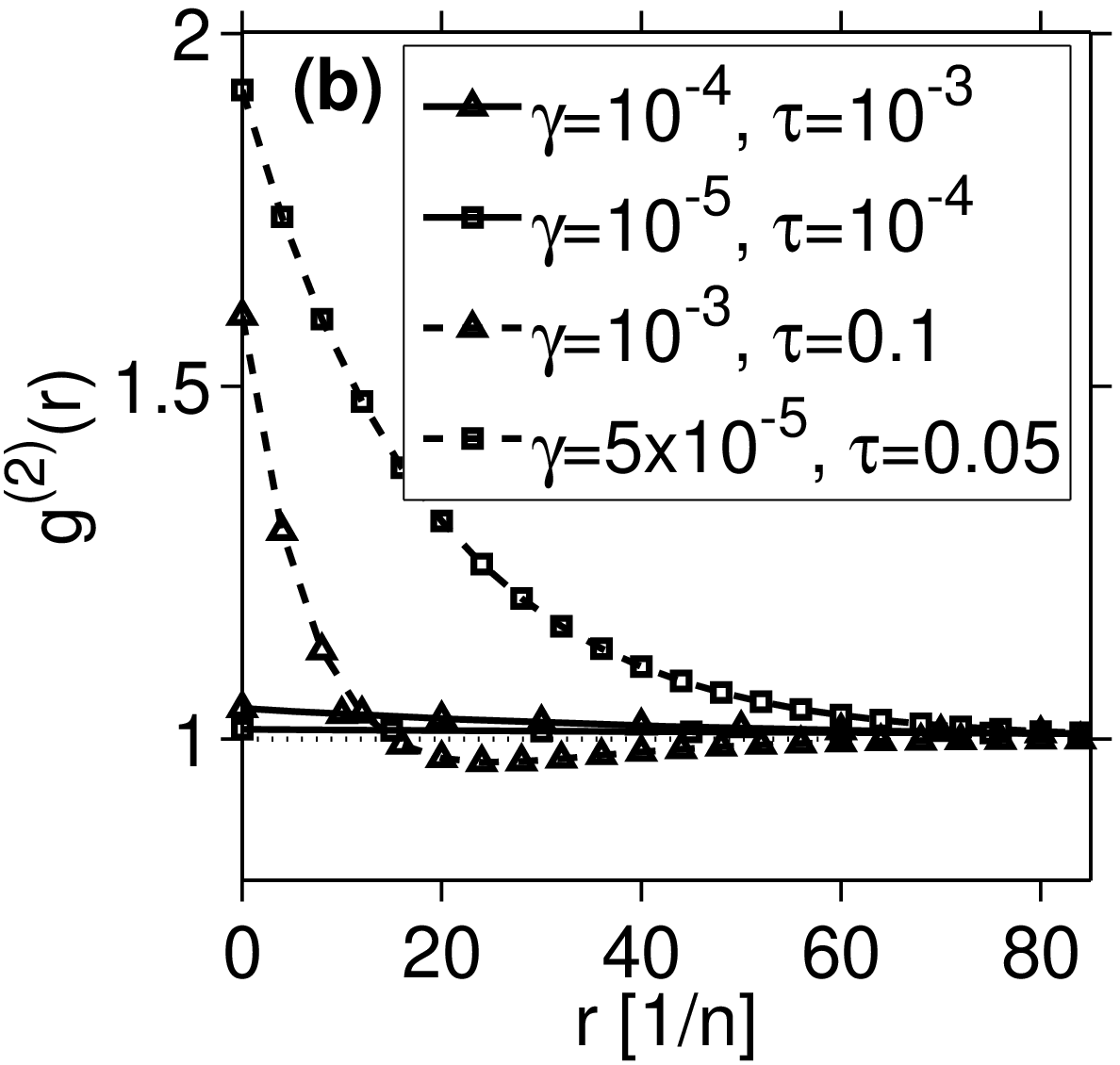}
\caption{Pair correlation $g^{(2)}(r)$ in the weakly interacting and nearly
ideal gas regimes. (a) Solid lines -- low-temperature weakly interacting gas
at $\protect\tau \ll \protect\gamma \ll 1$, Eq.~(\protect\ref{GPa}); dashed
lines -- DC regime, Eq.~(\protect\ref{DC}). (b) Solid lines -- weakly
interacting gas at $\protect\gamma \ll \protect\tau \ll \protect\sqrt{%
\protect\gamma }$, Eq.~(\protect\ref{GPb}); dashed lines -- DQ regime, Eq.~(%
\protect\ref{DQ}).}
\label{GPDQ}
\end{figure}

At sufficiently low temperatures, $\tau \ll \gamma \ll 1$, when
vacuum fluctuations dominate the excitations and thermal
fluctuations are a small correction we find \cite{g2r-epaps}
\begin{align}
g^{(2)}(r)& =1-\sqrt{\gamma }\left[ \mathbf{L}_{-1}(2\sqrt{\gamma }%
nr)-I_{1}(2\sqrt{\gamma }nr)\right]  \notag \\
& +\frac{1}{2\pi \sqrt{\gamma }n^{2}r^{2}}-\frac{\pi \tau ^{2}}{8\gamma
^{3/2}}\cosech^{2}\left( \frac{\tau n\pi r}{2\sqrt{\gamma }}\right) ,
\label{GPa}
\end{align}%
where $\mathbf{L}_{-1}(x)$ is the modified Struve function and $I_{1}(x)$ is
a Bessel function. The correlation length here is set by the healing length $%
\xi =\hbar /\sqrt{mgn}=1/\sqrt{\gamma }n$. For $r\gg \xi $ and finite $\tau $%
, the last term in Eq.~(\ref{GPa}) dominates the others and gives an
exponential decay to the uncorrelated value of $g^{(2)}(r)=1$ (for $\tau
\rightarrow 0$ one has a power law decay). Even at $T=0$, oscillating terms
are absent, in contrast to the strongly interacting regime, Eq.~(\ref{TGa}).
The limit $r\rightarrow 0$ reproduces the result of Eq.~(9) of Ref.~\cite%
{karen-prl}, $g^{(2)}(0)=1-2\sqrt{\gamma }/\pi +\pi \tau ^{2}/(24\gamma
^{3/2})$. In Fig.~\ref{GPDQ}(a) we plot Eq.~(\ref{GPa}) for different values
of $\gamma $, and we note that the finite temperature correction term is
negligible here.

In the opposite limit, dominated by thermal
fluctuations
corresponding to $\gamma \ll \tau \ll \sqrt{\gamma }$,
we find \cite{g2r-epaps}
\begin{align}
g^{(2)}(r)& =1+\frac{\tau }{2\sqrt{\gamma }}e^{-2\sqrt{\gamma }nr}  \notag \\
& -\sqrt{\gamma }\left[ \mathbf{L}_{-1}(2\sqrt{\gamma }nr)-I_{1}(2\sqrt{%
\gamma }nr)\right] ,  \label{GPb}
\end{align}%
valid for $r/\xi \alt1$. The last two terms are due to vacuum
fluctuations and are a negligible correction, so the leading term
gives an exponential decay [see Fig.~\ref{GPDQ}(b)] with a
correlation length given by the
healing length $\xi $. The peak value is $g^{(2)}(0)=1+\tau /(2\sqrt{\gamma }%
)$, in agreement with Ref.~\cite{karen-prl}. For $r/\xi \gg 1$ vacuum
fluctuations dominate and we reproduce the asymptotic behavior of Eq.~(\ref%
{GPa}).

\textit{Nearly ideal gas regime} [$\gamma \ll \min \{\tau
^{2},\sqrt{\tau }\} $]. --- Finally, we present the results for the
decoherent regime, where both the density and phase fluctuations are
large and the local pair correlation is close to the result for
noninteracting bosons, $g^{(2)}(0)=2$. Depending on the temperature
$\tau $, we further distinguish two sub-regimes: decoherent quantum
(DQ) regime for $\tau \ll 1$, and decoherent classical (DC) regime
for $\tau \gg 1$. Both can be treated using perturbation theory with
respect to the coupling constant $g$ around the ideal Bose gas
result of Ref.~\cite{Bouchoule-density-density}.

In the DQ regime, with $\sqrt{\gamma }\ll \tau \ll 1$ \cite{karen-prl}, the
nonlocal pair correlation is \cite{g2r-epaps}
\begin{equation}
g^{(2)}(r)=1+[1-4\gamma (1+\tau nr)/\tau ^{2}]e^{-\tau nr},  \label{DQ}
\end{equation}%
with the peak value $g^{(2)}(0)=2-4\gamma /\tau ^{2}$ \cite{karen-prl}. For $%
\gamma =0$ the correlations decay exponentially with the characteristic
correlation length which coincides with the phase coherence length $l_{\phi
}\simeq \hbar ^{2}n/mT=2/\tau n$ \cite{karen-prl} and is responsible for the
long-wavelength phase fluctuations. For $\gamma >0$, $g^{(2)}(r)$ becomes
nonmonotonic with a minimum at $nr_{\min }=\tau /4\gamma \gg 1$ before
reaching its uncorrelated value $g^{(2)}(r\rightarrow \infty )=1$. Thus, at
intermediate range we have weak antibunching due to interatomic repulsion,
while at short range we have typical ideal-gas bunching due to exchange
interaction, as shown in Fig.~\ref{GPDQ}(b).

In the DC regime ($\tau \gg \max \{1,\gamma ^{2}\}$), the pair correlation
is given by \cite{g2r-epaps}
\begin{equation}
g^{(2)}(r)=1+e^{-\tau n^{2}r^{2}/2}-\gamma \sqrt{2\pi /\tau }\;\mathrm{erfc}(
\sqrt{\tau n^{2}r^{2}/2}),  \label{DC}
\end{equation}
where $\mathrm{erfc}(x)$ is the complimentary error function. At $r=0$ we
have $g^{(2)}(0)=2-\gamma \sqrt{2\pi /\tau }$ \cite{karen-prl} as $\mathrm{
erfc}(0)=1$. In the noninteracting limit ($\gamma =0$) we recover the
well-known result for the classical ideal gas \cite{Naraschewski-Glauber}
characterized by Gaussian decay with a correlation length $\Lambda _{T}$.
For $\gamma > 0$ we observe [see Fig.~\ref{GPDQ}(a)] the emergence of
nonmonotonic behavior, with a global maximum $g^{(2)} (r_{\max}) = g^{(2)}
(0) + 2\gamma^2/\tau$ at nonzero separation $n r_{\max}=2\gamma/\tau \ll 1$.
As $\gamma$ is increased, there is a continuous transition from the DC
regime to the regime of high-temperature \textquotedblleft
fermionization\textquotedblright, with $g^{(2)}(0)$ reducing further and the
maximum moving to larger distances.

We now discuss experimental methods to measure pair correlations. Local
correlations have been measured by photoassociation~\cite{Weiss-g2-1D} and
by three-body loss~\cite{Phillips-g3-1997,Phillips-g3-1D-2003}, however,
nonlocal spatial correlations have not been measured \textit{in-situ} to the
best of our knowledge. We discuss one possible implementation of spatially
resolved imaging in the context of a 1D gas. The method is closely related
to the spatially resolved measurement of Ref.~\cite%
{thomas-spatially-resolved}. The first step is to \textquotedblleft
freeze-in\textquotedblright\ the correlations by turning on a deep standing
wave of far-detuned light, which breaks up the distribution into discrete
packets, separated by a half-wavelength.
The next step is to apply a magnetic field that varies linearly in
magnitude, creating a spatially dependent Zeeman shift. Next, a
stimulated Raman transition is driven such that atoms in two sites
are transferred to a different hyperfine state. This can be
accomplished by using two frequencies on the Raman beam
corresponding to two different locations. The unaffected atoms can
be removed with a pulse of resonant light
and the remaining atoms
counted.

For a specific example, consider a degenerate gas of sodium atoms in an
optical box trap~\cite{Raizen-BEC-box}, in the $F=2,m_{F}=-2$ state. After
\textquotedblleft freeze-in\textquotedblright\ with
a  $\lambda = 532$ nm
lattice, we apply a
magnetic field gradient of $150$~G/cm. We then drive a two-photon stimulated
Raman transition to the $F=1,m_{F}=-1$ state
with Raman beams at $532$ nm with optical power of $2.5$~mW focused to $50$~$\mu$m
and a pulse duration of $260$~$\mu$s. A detuning of $53$~kHz corresponds
to a shift of one lattice site and therefore a resolution of $\sim 266$ nm.
For typical 1D gas parameters, this resolution is sufficient to resolve the
characteristic correlation lengths found here. A related simpler method,
but with somewhat
lower resolution, has been recently proposed in Ref.
\cite{accordion}.

In summary, we have calculated nonlocal pair correlations in a uniform 1D
Bose gas in six physically relevant regimes. The correlations can be
measured using spatially resolved single-atom counting. We have shown
explicitly that the characteristic global correlation lengths are given by
one of four length scales: the thermal de Broglie wavelength, the mean
interparticle separation, the healing length, or the phase coherence length.
In all cases we identified the profound role of interactions that can lead
to non-trivial structures with local maxima or minima at a finite
interparticle separation.

AGS, MJD, and KVK acknowledge stimulating discussions with P. Deuar, P.
Drummond, A. Cherny and J. Brand, and the support by the Australian Research
Council. DMG acknowledges support by EPSRC Advanced Fellowship EP/D072514/1.
KVK and DMG acknowledge the hospitality and support of the Institut Henri
Poincar\'{e}. MGR acknowledges support from the R. A. Welch Foundation, The
National Science Foundation, and the Sid W. Richardson Foundation.

\end{document}